\documentclass[11pt]{article}
\usepackage{amsmath,amssymb,amsthm,amsxtra,overpic,bbm,bm,epsfig,subfigure,multirow}
\usepackage{color}

\usepackage{pdflscape}

\usepackage{comment}

\def\thefootnote{\fnsymbol{footnote}}

\begin{document}

\vspace{0.2cm}

\begin{center}
{\Large\bf Radiative Corrections to the Solar Lepton Mixing Sum Rule}
\end{center}

\vspace{0.2cm}

\begin{center}
{\bf Jue Zhang $^{a}$} \footnote{E-mail: zhangjue@ihep.ac.cn}
\quad {\bf Shun Zhou $^{a,~b}$} \footnote{E-mail: zhoush@ihep.ac.cn}
\\
{$^a$Institute of High Energy Physics, Chinese Academy of
Sciences, Beijing 100049, China \\
$^b$Center for High Energy Physics, Peking University, Beijing 100871, China}
\end{center}

\vspace{1.5cm}

\begin{abstract}

The simple correlation among three lepton flavor mixing angles $(\theta^{}_{12}, \theta^{}_{13}, \theta^{}_{23})$ and the leptonic Dirac CP-violating phase $\delta$ is conventionally called a sum rule of lepton flavor mixing, which may be derived from a class of neutrino mass models with flavor symmetries. In this paper, we consider the solar lepton mixing sum rule $\theta^{}_{12} \approx \theta^{\nu}_{12} + \theta^{}_{13} \cos \delta$, where $\theta^\nu_{12}$ stems from a constant mixing pattern in the neutrino sector and takes the value of $\theta^\nu_{12} = 45^\circ$ for the bi-maximal mixing (BM), $\theta^\nu_{12} = \tan^{-1}(1/\sqrt{2}) \approx 35.3^\circ$ for the tri-bimaximal mixing (TBM) or $\theta^\nu_{12} = \tan^{-1}\left[2/(\sqrt{5} + 1)\right] \approx 31.7^\circ$ for the golden-ratio mixing (GR), and investigate the renormalization-group (RG) running effects on lepton flavor mixing parameters when this sum rule is assumed at a superhigh-energy scale. For illustration, we work within the framework of the minimal supersymmetric standard model (MSSM), and implement the Bayesian approach to explore the posterior distribution of $\delta$ at the low-energy scale, which becomes quite broad when the RG running effects are significant. Moreover, we also discuss the compatibility of the above three mixing scenarios with current neutrino oscillation data, and observe that radiative corrections can increase such a compatibility for the BM scenario, resulting in a weaker preference for the TBM and GR ones.

\end{abstract}

\begin{flushleft}
\hspace{0.8cm} PACS number(s): 11.10.Hi, 14.60.Pq
\end{flushleft}

\def\thefootnote{\arabic{footnote}}
\setcounter{footnote}{0}

\newpage

\section{Introduction}

Thanks to the dedicated experimental efforts in the last two decades, our knowledge on neutrinos has been greatly improved. It is now a well-established fact that neutrinos are massive and three lepton flavors are significantly mixed~\cite{Agashe:2014kda}. The ongoing and forthcoming experiments will further unravel the mysteries of neutrinos, such as the neutrino mass ordering, the size of CP violation in the lepton sector, the absolute neutrino mass scale and the Majorana or Dirac nature of neutrinos (i.e., whether neutrinos are their own antiparticles). On the other hand, future neutrino oscillation experiments can also measure the currently known mixing parameters to a higher precision level.

The precision measurements of neutrino mixing parameters will provide us with a great opportunity to test the neutrino mass models that account for both tiny neutrino masses and large lepton flavor mixing. Although the theoretical predictions from neutrino mass models are often model-dependent, there actually exist some generic model-independent ones. One example of these model-independent predictions is the sum rule that imposes a relation among absolute neutrino masses~\cite{mass_sum_rule_1, Barry:2010yk, Dorame:2011eb, King:2013psa, Gehrlein:2015ena} or neutrino mixing parameters~\cite{SSR_early, Haba:2012qx, King:2013eh, Ballett:2013wya, Petcov:2014laa, Girardi:2014faa, Ballett:2014dua, Girardi:2015rwa}. Given the fact that these sum rules are usually derived from neutrino mass models with flavor symmetries that are supposed to work at a superhigh-energy scale, one then inevitably needs to take into account the corrections to neutrino masses and mixing parameters from the RG running. For instance, in a recent work~\cite{Gehrlein:2015ena}, the RG running effects on the sum rule of neutrino masses have been investigated.

In this paper, we study the impact of RG running effects on the sum rule of lepton flavor mixing. As is well known, the RG running effects on the lepton mixing parameters are insignificant in the Standard Model (SM), so the sum rule can be directly confronted with neutrino oscillation data. We therefore consider scenarios that go beyond the SM, and for illustration we choose to work within the framework of MSSM, as it is known that large RG running effects on lepton mixing parameters can be present if $\tan\beta$ is relatively large. Moreover, since in MSSM the mixing angle $\theta^{}_{12}$ is much more sensitive to the running effects than the other two mixing angles $\theta^{}_{13}$ and $\theta^{}_{23}$~\cite{Antusch:2003kp,Antusch:2005gp,Mei:2005qp,Mei:2005gp,Luo:2005fc,Ohlsson:2013xva}, we then examine only the sum rule for $\theta^{}_{12}$ or the so-called solar mixing sum rule, which can be derived from a general class of flavor symmetry models. To the leading order, such a sum rule is approximately given as $\theta_{12} \approx \theta_{12}^\nu + \theta_{13}^{} \cos\delta$~\cite{SSR_early}, where $\theta^\nu_{12}$ stems from a constant mixing pattern in the neutrino sector and takes the value of $\theta^\nu_{12} = 45^\circ$ for the BM mixing~\cite{BM}, $\theta^\nu_{12} = \tan^{-1}(1/\sqrt{2}) \approx 35.3^\circ$ for the TBM mixing~\cite{TBM} or $\theta^\nu_{12} = \tan^{-1}\left[2/(\sqrt{5} + 1)\right] \approx 31.7^\circ$ for the GR mixing~\cite{GR_A}.\footnote{There exist alternative mixing patterns involving the golden ratio, which assume $\theta^\nu_{12} = \cos^{-1}\left[(\sqrt{5} + 1)/4\right] = 36.0^\circ$~\cite{GR_B}, or $\theta^\nu_{12} = \tan^{-1}\left[2/(\sqrt{5} + 3)\right] = 20.9^\circ$~\cite{GR_C}. In the former case, the resultant solar mixing sum rule would be quite similar to that for the TBM mixing, which we shall study in detail in this work. In the latter case, significant radiative corrections to the corresponding sum rule are needed to reconcile theoretical predictions of mixing parameters with low-energy neutrino oscillation data. This kind of leptonic mixing sum rule deserves further studies in a future publication.} According to the latest global-fit analysis of neutrino oscillation data, as shown in Table~\ref{tb:fit}, one can observe that imposing the aforementioned sum rule directly at the low-energy scale would yield a prediction of $\delta \approx 180^\circ$ for the BM mixing, while $\delta \approx 90^\circ$ or $270^\circ$ for the TBM and GR mixings. These predictions are valid in the case when RG corrections are insignificant, so a future precise measurement of $\delta$ can be used to discriminate BM from the other two. The further discrimination between TBM and GR would require an even higher precision, which may not be reached in a near future.

Our study is different from Ref.~\cite{Gehrlein:2015ena} and other relevant works~\cite{Ballett:2014dua} in several aspects. First, only the sum rule among three mixing angles $(\theta^{}_{12}, \theta^{}_{13}, \theta^{}_{23})$ and the Dirac CP-violating phase $\delta$ is considered, but neutrino masses $(m^{}_1, m^{}_2, m^{}_3)$ and the Majorana CP-violating phases $(\varphi^{}_1, \varphi^{}_2)$ are set to be free. Second, a particular attention is given to the unknown CP-violating phase $\delta$. We investigate how the RG running effects modify the predicted values of $\delta$ from the solar mixing sum rule that is valid at a superhigh-energy scale. Quantitatively, the Bayesian statistical approach is adopted to calculate the posterior distribution of $\delta$ at the low-energy scale. We find that the prediction for $\delta$ is quite sensitive to RG corrections; the modification can be of order $\mathcal{O}(10^\circ)$, or even $180^\circ$ in some extreme cases. Such modifications would complicate the test on a certain solar mixing sum rule, and also blur the discrimination among different solar mixing sum rules. Finally, the compatibility of the sum rule corresponding to the BM, TBM or GR mixing with neutrino oscillation data is examined by including the running effects. In order to fully test these mixing sum rules, one needs to first find out if the running effects are important or not. This might be achieved in the future by pinning down the neutrino mass ordering and the absolute neutrino mass scale, and by constraining the value of $\tan\beta$ in the MSSM. It should be noted that the above findings are restricted to the scenario of MSSM. For other possible extensions of the SM, one may have to study the RG running effects in question separately, as both the corresponding RG equations of lepton mixing parameters and the energy range available for RG running can be quite different. Taking the universal extra-dimensional models~\cite{extra_dim} for example, in which all the SM fields are allowed to propagate in one or more compact extra dimensions, one can find that the running of neutrino parameters will obey a power law due to the increasing number of excited Kaluza-Klein modes, implying a remarkable boost in the running within a relatively narrow range of energy scales~\cite{RG_extra_dim}.

The remaining part of our paper is organized as follows. In Section 2, we provide with a quick overview on the solar mixing sum rule, and derive formulas that account for the RG corrections. We then introduce our numerical method in Section 3, followed by a detailed discussion on four examples in Section 4.  In Section 5, we present our global numerical results for all the cases that have been studied in this work. Lastly, we summarize our results in Section 6.

\begin{table}[t]
\begin{center}
\vspace{-0.25cm} \caption{The best-fit values, together with the
1$\sigma$, 2$\sigma$ and 3$\sigma$ intervals, for three neutrino
mixing angles $\{\theta^{}_{12}, \theta^{}_{13}, \theta^{}_{23}\}$, two mass-squared differences $\{\Delta m^2_{21} \equiv m^2_2 - m^2_1, \Delta m^2_{31}\equiv m^2_3 - m^2_1~{\rm or}~\Delta m^2_{32} \equiv m^2_3 - m^2_2\}$ and the Dirac CP-violating phase $\delta$ from a global analysis of current experimental data~\cite{Gonzalez-Garcia}. Several independent global-fit analyses can be found in Refs.~\cite{Fogli,Valle,Bergstrom:2015rba}, which are in perfect agreement with the results presented here at the $3\sigma$ level.} \vspace{0.2cm}
\label{tb:fit}
\begin{tabular}{c|c|c|c|c}
\hline
\hline
Parameter & Best fit & 1$\sigma$ range & 2$\sigma$ range & 3$\sigma$ range \\
\hline
\multicolumn{5}{c}{Normal neutrino mass ordering (NO)
$(m^{}_1 < m^{}_2 < m^{}_3$)} \\ \hline
$\theta_{12}/^\circ$
& $33.48$ & 32.73 --- 34.26 & 31.98 --- 35.04 & 31.29 --- 35.91 \\
$\theta_{13}/^\circ$
& $8.50$ & 8.29 --- 8.70 & 8.08 --- 8.90 & 7.85 --- 9.10 \\
$\theta_{23}/^\circ$
& $42.3$  & 40.7 --- 45.3 & 39.1 --- 48.3 & 38.2 --- 53.3 \\
$\delta/^\circ$ &  $306$ & 236 --- 345 & 0 --- 24
$\oplus$ 166 --- 360 & 0 --- 360 \\
$\Delta m^2_{21}/[10^{-5}~{\rm eV}^2]$ &  $7.50$ & 7.33 --- 7.69 & 7.16 --- 7.88 & 7.02 --- 8.09 \\
$\Delta m^2_{31}/[10^{-3}~{\rm eV}^2]$ &  $+2.457$ & +2.410 --- +2.504 & +2.363 --- +2.551 & +2.317 --- +2.607 \\\hline
\multicolumn{5}{c}{Inverted neutrino mass ordering (IO)
$(m^{}_3 < m^{}_1 < m^{}_2$)} \\ \hline
$\theta_{12}/^\circ$
& $33.48$ & 32.73 --- 34.26 & 31.98 --- 35.04 & 31.29 --- 35.91 \\
$\theta_{13}/^\circ$
& $8.51$ & 8.30 --- 8.71 & 8.09 --- 8.91 & 7.87 --- 9.11 \\
$\theta_{23}/^\circ$
& $49.5$  & 47.3 --- 51.0 & 45.1 --- 52.5 & 38.6 --- 53.3 \\
$\delta/^\circ$ &  $254$ & 192 --- 317 & 0 --- 20
$\oplus$ 130 --- 360 & 0 --- 360 \\
$\Delta m^2_{21}/[10^{-5}~{\rm eV}^2]$ &  $7.50$ & 7.33 --- 7.69 & 7.16 --- 7.88 & 7.02 --- 8.09 \\
$\Delta m^2_{32}/[10^{-3}~{\rm eV}^2]$ &  $-2.449$ & $-2.496$ --- $-2.401$ & $-2.543$ --- $-2.355$ & $-2.590$ --- $-2.307$ \\ \hline\hline
\end{tabular}
\end{center}
\end{table}

\section{Solar mixing sum rule and RG corrections}

\subsection{Solar mixing sum rule}

Given the fact that neutrinos are massive, the lepton flavor mixing~\cite{Maki:1962mu,Pontecorvo:1957cp} is described by the Maki-Nakagawa-Sakata-Pontecorvo (MNSP) matrix $U =U_{l}^{\dagger} U_\nu^{}$, where $U_{l}^{}$ and $U_\nu^{}$ are the unitary matrices that arise from the diagonalization of the charged-lepton mass matrix and the neutrino mass matrix, respectively. Assuming that neutrinos are Majorana particles, we have the standard parameterization
\begin{eqnarray} \label{eq:MNSP}
U = V(\theta_{12}^{}, \theta_{13}^{}, \theta_{23}^{}, \delta) \cdot \text{Diag}(e^{-i\varphi_1/2}_{}, e^{-i\varphi_2/2}_{}, 1) \; ,
\end{eqnarray}
with
\begin{eqnarray}
V = \begin{pmatrix}
c_{12}^{} c_{13}^{} & s_{12}^{} c_{13}^{} &  s_{13}^{}e^{-i\delta}_{} \\
-s_{12}^{} c_{23}^{} - c_{12}^{} s_{13}^{} s_{23}^{} e^{i\delta}_{} & c_{12}^{} c_{23}^{} - s_{12}^{} s_{13}^{} s_{23}^{} e^{i\delta}_{} & c_{13}^{} s_{23}^{} \\
s_{12}^{} s_{23}^{} - c_{12}^{} s_{13}^{} c_{23}^{} e^{i\delta}_{} & - c_{12}^{} s_{23}^{} - s_{12}^{} s_{13}^{} c_{23}^{} e^{i\delta}_{} & c_{13}^{} c_{23}^{}
\end{pmatrix},
\end{eqnarray}
where $c^{}_{ij} \equiv \cos \theta^{}_{ij}$ and $s^{}_{ij} \equiv \sin \theta^{}_{ij}$ (for $ij = 12, 13, 23$) have been defined, and $\delta$ and $\varphi_{1,2}^{}$ are the Dirac and Majorana CP-violating phases, respectively.

In the neutrino mass model with a discrete flavor symmetry (see, e.g., Ref.~\cite{King:2013eh} for a recent review), it is quite common that the unitary matrix $U^{}_\nu$ arising from the diagonalization of neutrino mass matrix takes a particular form of the BM, TBM or GR mixing, which implies $\theta_{23}^\nu = 45^\circ$ and $\theta_{13}^\nu = 0$ when the standard parametrization as in Eqs.~(1) and (2) is applied to $U^{}_\nu(\theta^\nu_{12}, \theta^\nu_{13}, \theta^\nu_{23})$. In this case, $U_\nu^{}$ can be in general parametrized by
\begin{eqnarray}
U_\nu^{} = P_{\rm L}^\nu \begin{pmatrix}
c_{12}^\nu & s_{12}^\nu & 0 \\
-s_{12}^\nu/\sqrt{2} & c_{12}^\nu/\sqrt{2} & 1/\sqrt{2} \\
s_{12}^\nu/\sqrt{2} & -c_{12}^\nu/\sqrt{2} & 1/\sqrt{2}
\end{pmatrix} P_{\rm R}^\nu \; ,
\end{eqnarray}
where $P_{\rm L,R}^\nu$ are diagonal phase matrices, and $c^\nu_{12} \equiv \cos \theta^\nu_{12}$ and $s^\nu_{12} \equiv \sin \theta^\nu_{12}$ are implied. The phases in $P_{\rm L}^\nu$ are to be combined with those from $U^{}_l$ and will contribute to the final MNSP matrix. On the other hand, the phases in $P_{\rm R}^\nu$ simply drop out in the final stage of extracting mixing angles and the Dirac CP-violating phase from $U$, but they are relevant for the Majorana CP-violating phases.

Since a non-zero value of $\theta^{}_{13}$ has been discovered~\cite{An:2012eh}, $U_\nu^{}$ alone is unable to describe the measured lepton flavor mixing angles and small corrections in $U^{}_l$ from the charged-lepton sector are needed. Motivated by the observation that the charged-lepton masses exhibit a very strong hierarchy as quark masses do, one can make a further assumption that the rotation angle $\theta_{13}^l$ in $U_l^{}$ is vanishingly small. Consequently, as shown in Ref.~\cite{Ballett:2014dua}, there exists a simple but instructive relation for the matrix elements of $U$, namely, $|U^{}_{\tau 1}| / |U^{}_{\tau 2}| = \tan \theta_{12}^\nu$, where $U_{\tau 1}$ and $U_{\tau 2}$ are the first two elements in the last row of the MNSP matrix. With the standard parameterization of $U$ given in Eq.~(\ref{eq:MNSP}), we then obtain the exact form of the solar mixing sum rule~\cite{Petcov:2014laa, Ballett:2014dua}
\begin{eqnarray} \label{eq:SSR}
\cos\delta = \frac{(s_{12}^2 - s_{12}^{\nu 2}) t_{23}^{}}{2 s_{12}^{} c_{12}^{} s_{13}^{}} - \frac{(s_{12}^2 -c_{12}^{\nu 2})  s_{13}^{}}{2s_{12}^{} c_{12}^{} t_{23}^{}} \; ,
\end{eqnarray}
with $t^{}_{23} \equiv \tan \theta^{}_{23}$. Noticing that $s_{13}^{} \sim |\theta_{23}^{} - \pi/4| \sim |\theta_{12}^{} -\theta_{12}^\nu| \ll 1$ holds for the BM, TBM and GR constant mixing patterns, one can reduce Eq.~(\ref{eq:SSR}) to a much simpler form $\theta_{12} \approx \theta_{12}^{\nu} + \theta_{13}\cos\delta$, which characterizes the deviation of $\theta^{}_{12}$ from $\theta^\nu_{12}$ by a combination of the smallest mixing angle $\theta^{}_{13}$ and the yet-unknown CP-violating phase $\delta$. For the BM mixing with $\theta^\nu_{12} = 45^\circ$, which is larger than the best-fit value $\theta^{}_{12} = 33.48^\circ$ in Table~\ref{tb:fit}, $\delta \approx 180^\circ$ is preferred to relax the tension between the solar mixing sum rule and neutrino oscillation data. For the TBM mixing with $\theta^\nu_{12} \approx 35.3^\circ$ and the GR mixing with $\theta^\nu_{12} \approx 31.7^\circ$, which are already consistent with the observed $\theta^{}_{12}$ within $2\sigma$, a maximal CP-violating phase $\delta \approx 90^\circ$ or $270^\circ$ is implied by the sum rule. Although the leading-order form of the solar mixing sum rule appears simple, we shall employ the exact expression in Eq.~(\ref{eq:SSR}) when performing analytical and numerical studies on the RG corrections, as significant running effects can spoil the above approximations. Note that an alternative derivation of the above sum rule can be found in Ref.~\cite{Girardi:2014faa}, where individual mixing matrices in $U$ are carefully relocated, and eventually one realizes that $U$ can be described by only three free parameters after separating out the Majorana CP-violating phases. Therefore, one sum rule among the three mixing angles and the Dirac CP-violating phase is obtained.

Before discussing the RG corrections to the solar mixing sum rule, we briefly summarize the previous results without RG running effects. In Ref.~\cite{Ballett:2014dua}, one assumes some prior distributions that are compatible with the latest global-fit results of three mixing angles, and then predicts the posterior distribution of the Dirac CP-violating phase according to Eq.~(\ref{eq:SSR}). It has been found that the predicted value of $\cos\delta$ in the BM case is centered far below $-1$, with only a tiny tail above it, while for TBM and GR the prediction of $\cos\delta$ is well within the range of $[-1, 1]$, and their central values are close to 0. More explicitly, if taking the best-fit values of all the three mixing angles in the case of NO from Table \ref{tb:fit}, we find $\cos\delta = -1.27$ for BM, while $\cos\delta = - 0.13$ and $\cos\delta = 0.27$ for TBM and GR, respectively. Thus, if the RG corrections are negligible, the scenario of BM mixing is already disfavored by current data, while TBM and GR are still allowed. An important motivation of this work is to investigate whether such a conclusion still holds when RG running effects are considered.

\subsection{RG corrections to solar mixing sum rule}

In this subsection, we perform an analytical study on the RG corrections to the solar mixing sum rule. For later convenience, let us introduce a parameter $\Delta$ to describe how much the sum rule is violated, namely,
\begin{eqnarray} \label{eq:Delta}
\Delta \equiv 2s_{12}^{} c_{12}^{} s_{13}^{} \cos\delta  + (s_{12}^{\nu 2} - s_{12}^2) t_{23}^{} + (s_{12}^2 - c_{12}^{\nu 2})s_{13}^{2} / t_{23}^{} \; ,
\end{eqnarray}
from which one can verify that Eq.~(\ref{eq:SSR}) corresponds to $\Delta = 0$. Note that the mixing angles, CP-violating phases and $\Delta$ depend actually on the renormalization scale~\cite{Ohlsson:2013xva}. If the solar mixing sum rule is first derived at a superhigh-energy scale, we have $\Delta = 0$ but it may become non-zero at the low-energy scale due to radiative corrections. In order to avoid any confusion, we denote the RG-corrected value of $\Delta$ at low energies by $\Delta^{\rm L}$. Hence, one can predict the Dirac CP-violating phase $\delta^{\rm L}$ at low energies in terms of the low energy mixing angles $\{\theta_{12}^{\rm L}, \theta_{13}^{\rm L}, \theta_{23}^{\rm L}\}$ by using Eq.~(\ref{eq:Delta}). Note that here any quantity with the superscript ``L" stands for its low-energy value that is derived from the RG evolution of its initial value at the high-energy scale, at which the sum rule is satisfied.

The evolution of $\Delta$ can be found by solving the RG equations of three mixing angles and CP-violating phases, which have been summarized in Appendix A. Expanding the RG equations in terms of $\theta^{}_{13}$, one can find that $\dot{\theta}_{ij} \sim \mathcal{O}(\theta_{13}^{0})$ and $\dot{\delta} \sim \mathcal{O}(\theta_{13}^{-1})$, where $\dot{f} = df/dt$ with $t = \ln \mu$ and $\mu$ is the renormalization scale. Therefore, the RG evolution of $\Delta$ is governed by the following equation
\begin{eqnarray} \label{eq:dot_Delta}
\dot{\Delta} = -2 (\sin\delta) s_{12}^{} c_{12}^{} s_{13}^{} \dot{\delta} -2 s_{12}^{} c_{12}^{} t_{23}^{} \dot{\theta}_{12}^{} + 2(\cos\delta)s_{12}^{} c_{12}^{} \dot{\theta}_{13}^{} + (s_{12}^{\nu 2}-s_{12}^{2})/c_{23}^2  \dot{\theta}_{23} + \mathcal{O}(s_{13}^{}) \; ,
\end{eqnarray}
which is the master equation of this work. Based on Eq.~(\ref{eq:dot_Delta}), we discuss two extreme scenarios for small and large RG running effects, respectively.
\begin{itemize}
\item \emph{Small running effects} -- This scenario can arise when three neutrino masses are hierarchical, or when the value of $\tan\beta$ in the MSSM is relatively small. Now that all the mixing angles and CP-violating phases do not run significantly in this case, we can integrate the evolution equation of $\Delta$ by assuming the coefficient of each term on the right-hand side of Eq.~(\ref{eq:dot_Delta}) to be constant. Denoting the RG correction to a parameter $f$ as $\Delta f$, we obtain
\begin{eqnarray}
\cos\delta^{\rm L} \approx \cos\delta_0 - \sin\delta^{\rm L} (\Delta\delta) - t_{23}^{\rm L} \left(  \frac{\Delta \theta_{12}^{}}{\theta_{13}^{\rm L}} \right) + \cos\delta^{\rm L} \left( \frac{\Delta \theta_{13}^{}}{\theta_{13}^{\rm L}} \right) +  \frac{s_{12}^{\nu 2}-s_{12}^{\rm L 2}}{2 s_{12}^{\rm L} c_{12}^{\rm L} c_{23}^{\rm L 2}}  \left( \frac{\Delta \theta_{23}^{}}{\theta_{13}^{\rm L}} \right),
\end{eqnarray}
where $\cos \delta^{}_0$ is given by Eq.~(\ref{eq:SSR}) but with the mixing angles on the right-hand side substituted by their RG-corrected values $\theta^{\rm L}_{ij}$ (for $ij = 12, 13, 23$). It is then clear that when $\Delta\theta_{ij}^{}$ is comparable to $\theta_{13}^{\rm L}$ significant modifications to the Dirac CP-violating phase $\delta$ can arise.

\item \emph{Nearly-degenerate mass spectrum} -- As is well-known, all the mixing angles and CP-violating phases can receive remarkable RG running effects~\cite{Antusch:2003kp} when the lightest neutrino mass $m_0^{}$ is relatively large, i.e., $m_0^{} \gtrsim 0.05~{\rm eV}$ and neutrino mass spectrum is nearly degenerate. Assuming three light neutrino masses $m_i^{} \approx m_0^{}$ (for $i = 1, 2, 3$), we find that the RG equation of $\Delta$ is approximately given by
\begin{eqnarray} \label{eq:dot_Delta_deg}
\frac{32\pi^2}{y_\tau^2} \dot{\Delta} &\approx & \sin^2(2\theta_{12}^{})\sin(2\theta_{23}^{}) \frac{m_0^2}{\Delta m_{32}^2} (\cos\varphi_1^{} - \cos\varphi_2^{}) \nonumber \\
& & -~ 4 t_{23}^{} ( s_{12}^{\nu 2} - s_{12}^2) \frac{m_0^2}{\Delta m_{32}^2} ( 1 + c_{12}^2 \cos\varphi_2^{} + s_{12}^2 \cos\varphi_1^{}) \nonumber \\
& & + 2 \sin^2(2\theta_{12}) t_{23}^{} s_{23}^2 \frac{m_0^2}{\Delta m_{21}^2} [1 + \cos(\varphi_2^{} -\varphi_1^{})] \; ,
\end{eqnarray}
where $y_\tau^{}$ is the tau-lepton Yukawa coupling. Given $|\Delta m_{32}^2| \gg \Delta m_{21}^2$, then if $(\varphi_2^{} -\varphi_1^{})$ is far away from $180^\circ$, the dominant contribution to $\dot{\Delta}$ comes from the last term due to the enhancement from $m_0^2/\Delta m_{21}^2$. Moreover, such a contribution is always positive, resulting in a more negative value of $\Delta$ (or $\cos\delta$) at low energies. As a consequence, the BM case would become even more incompatible with the low energy data, while for TBM and GR the predicted values of $\delta^{\rm L}$ tend to be in the second or third quadrant. In fact, because of this observation, we will see in the later numerical studies that the relation $(\varphi_2^{} -\varphi_1^{}) \sim 180^\circ$ arises in the case of BM, and even in TBM and GR when RG runnings effects are sizeable.
\end{itemize}

After a brief discussion about two extreme scenarios in an analytical way, we then turn to the numerical study, where a more detailed discussion on the RG running effects will be given.

\section{Numerical method}

Our method for numerical studies is to adopt the notion of Bayesian statistical analysis, by focusing on the posterior distribution of $\delta$ at low energies. Before presenting the detailed numerical procedure, let us first discuss the application of general Bayesian analysis formalism to our work. See, e.g., Refs.~\cite{SSbook,Cowan:1998ji}, for more details about Bayesian analysis.


The Bayesian analysis resides in the well-known Bayes' theorem ${\rm Pr}(A|B) = {\rm Pr}(B|A) {\rm P}(A)/{\rm Pr}(B)$, where ${\rm Pr}(B|A)$ is the probability of the proposition $B$ to be true under the condition that $A$ is true, and likewise for ${\rm Pr}(A|B)$. In this work we are interested in the question: given the current neutrino data, what the consequences on the neutrino parameters are when further imposing the solar sum rule at some high energy scale. Thus, we take $A$ as parameters of interest in this work, namely, the lepton mixing angles and CP-violating phases, and they are collectively denoted as $\Theta$. On the other hand, $B$ represents one of our hypotheses $\mathcal{H}$, the satisfaction of a particular solar sum rule at high energy. With these identifications, we then rewrite the above theorem as
\begin{eqnarray}
{\rm Pr}(\Theta|\mathcal{H}) = \frac{{\rm Pr}(\mathcal{H}|\Theta) {\rm Pr}(\Theta)}{{\rm Pr}(\mathcal{H})}.
\end{eqnarray}
Here ${\rm Pr}(\Theta|\mathcal{H})$ denotes the posterior probability distribution of neutrino parameters assuming the satisfaction of sum rule at high energy, while ${\rm Pr}(\mathcal{H}|\Theta)$ is the likelihood function of satisfying the solar sum rule at high energy given a particular set of neutrino parameters. Furthermore, ${\rm Pr}(\Theta)$ is the prior probability distribution of neutrino parameters. Since current neutrino data have been taken as the background information, $\Theta$ is then taken to be lepton mixing parameters that are given at \emph{low} energies, and their prior distribution ${\rm Pr}(\Theta)$ will be determined according to the latest global-fit analysis of neutrino oscillation data. Lastly, we identify ${\rm Pr}(\mathcal{H})$ as the probability of satisfying the solar sum rule at high energy in face of the current data. Since the posterior distribution needs to be normalized, i.e., $\int \mathrm{Pr}(\Theta|\mathcal{H})~ \mathrm{d}^N\Theta = 1$, where $N$ is the dimension of the set of free parameters $\Theta$, we then have
\begin{eqnarray}
{\rm Pr}(\mathcal{H}) =  \int \mathrm{Pr}(\mathcal{H}|\Theta) ~\mathrm{Pr}(\Theta)~ \mathrm{d}^N\Theta.
\end{eqnarray}
Note that when obtaining the posterior probability distribution for neutrino parameters, this ${\rm Pr}(\mathcal{H})$ is a common factor for all $\Theta$, and therefore it is often ignored. However, in the later comparison of several different hypotheses it will play an important role.

The RG evolution of lepton mixing parameters has been extensively studied in the literature, and recently summarized in Ref.~\cite{Ohlsson:2013xva}. Our strategy in this work can be described as follows:
\begin{itemize}
\item In the flavor basis where the charged-lepton mass matrix $M^{}_l$ is diagonal, lepton flavor mixing and neutrino masses are determined by the effective neutrino mass matrix $M^{}_\nu$, arising from the dimension-five Weinberg operator~\cite{Weinberg:1979sa} after the spontaneous breakdown of electroweak gauge symmetry. In the seesaw models, the Weinberg operator emerges naturally below the seesaw scale, where the heavy degrees of freedom have been integrated out. Therefore, one can calculate the radiative corrections to lepton flavor mixing parameters by investigating the RG evolution of the effective neutrino mass matrix $M^{}_\nu$ in the framework of effective theories~\cite{Chankowski:1993tx}.

\item In our calculations, one-loop RG equations for the effective neutrino mass matrix, gauge couplings and Yukawa couplings in the MSSM are adopted. The RG running is chosen to start from the energy scale of 1 TeV, at which we input the values of various gauge and Yukawa couplings from Ref.~\cite{Antusch:2013jca}.\footnote{For simplicity, we do not include the supersymmetric threshold corrections~\cite{SUSY}. Adding them would make the numerical results less tractable, so we defer it to a future work.} To reconstruct the neutrino mass matrix, we assume Gaussian priors for the sines of lepton flavor mixing angles and two neutrino mass-squared differences according to the recent global-fit result~\cite{Gonzalez-Garcia}, while uniform priors on $[0, 360^\circ]$ are used for the CP-violating phases. For the lightest neutrino mass $m_0^{}$, we take four reference values, i.e., $m_0^{} =0.005~{\rm eV}$, $0.015~{\rm eV}$, $0.05~{\rm eV}$ and $0.15~\text{eV}$ at low energies.\footnote{Although $m_0^{} = 0.15 ~\text{eV}$ is already disfavored by the Planck 2015 result~\cite{Ade:2015xua}, i.e., $\sum m_\nu^{} < 0.23 ~\text{eV}$, we still adopt it for two reasons: first, this large value may still be possible if some assumptions in the Planck analysis are relaxed; second, we can employ it to represent the scenario where RG running effects are violent.} The value of $\tan\beta$ is also crucial, so we consider three benchmark values $\tan\beta = 10$, $30$ and $50$. The input parameters can be found in Table~\ref{tb:input}.

\begin{table}[t]
\centering
\vspace{-0.25cm} \caption{Details of the input parameters at low energies and the values of $\tan\beta$. Here $\mathcal{N}(\mu, \sigma)$ stands for a Gaussian distribution with mean $\mu$ and standard deviation $\sigma$, while the uniform prior on $[a, b]$ is denoted by $\mathcal{U}(a, b)$. Numerical values in each Gaussian prior are obtained by taking the best fit value in  Ref.~\cite{Gonzalez-Garcia} as mean $\mu$ and symmetrizing the corresponding $1\sigma$ errors for standard deviation $\sigma$.} \vspace{0.2cm}
\begin{tabular}{c | c | c}
\hline
\hline
 & NO & IO \\
\hline
$\sin^2\theta_{12}^{}$ & $\mathcal{N}(0.304, ~0.0125)$ & $\mathcal{N}(0.304, ~0.0125)$ \\
$\sin^2\theta_{13}^{}$ & $\mathcal{N}(0.0218, ~0.001)$ & $\mathcal{N}(0.0219, ~0.001)$ \\
$\sin^2\theta_{23}^{}$ & $\mathcal{N}(0.452, ~0.04)$ & $\mathcal{N}(0.579, ~0.031)$ \\
$\Delta m_{\rm 21}^2/ [10^{-5}~{\rm eV}^2]$ & $\mathcal{N}(7.50, ~0.18)$ & $\mathcal{N}(7.50, ~0.18)$ \\
$|\Delta m_{\rm 31}^2|/ [10^{-3}~{\rm eV}^2]$ & $\mathcal{N}(2.457, ~0.047)$ & $\mathcal{N}(2.449, ~0.0475)$ \\
\hline
$\delta, \varphi_{1,2}$ & \multicolumn{2}{c}{$\mathcal{U}(0,~ 360^\circ)$} \\
$m_0^{}/[{\rm eV}]$ & \multicolumn{2}{c}{0.005, 0.015, 0.05, 0.15} \\
\hline
$\tan\beta$ & \multicolumn{2}{c}{10, 30, 50} \\
\hline
\hline
\end{tabular}
\label{tb:input}
\end{table}

\item With low energy boundary values specified, we then run all the physical parameters via one-loop RG equations to a superhigh-energy scale, which will be fixed as $\Lambda = 10^{12}~{\rm GeV}$. Since RG running depends logarithmically on the energy scale, varying this high energy boundary scale by one order of magnitude should not change our results significantly. At the high-energy scale, we impose the exact form of the solar mixing sum rule by defining the likelihood function
\begin{eqnarray}
{\rm Pr}(\mathcal{H}|\Theta) \propto e^{-\chi^2/2}\quad {\rm with} \quad \chi^2 \equiv \left( \frac{\Delta^{\rm H} -0}{\sigma_\Delta^{}} \right)^2,
\end{eqnarray}
where $\Delta^{\rm H}_{}$ is the value of $\Delta$ at high energies, and the size of $\sigma_{\Delta}^{}$ characterizes the tolerance of satisfying the sum rule. Here we choose a very small value for $\sigma_{\Delta}^{}$, i.e., $\sigma_{\Delta}^{} = 0.0001$, to ensure the satisfaction of the sum rule.
\end{itemize}

Finally, numerical results in this work are obtained with the help of the M{\scriptsize ULTI}N{\scriptsize EST} program~\cite{multinest}, which not only evaluates ${\rm Pr}(\mathcal{H})$ efficiently but also generates the posterior distributions as a by-product.

\section{Four illustrative examples}

In this section, we present four illustrative examples to understand the RG corrections to the mixing sum rule. Since the predictions for $\theta^\nu_{12}$ are similar in the TBM and GR cases, we consider two examples for BM and another two for TBM, and the discussions on TBM can be readily applied to GR. Furthermore, only the scenario of NO is studied. The scenario of IO can also be similarly studied, although because of more intense RG corrections in IO some cases are less tractable, and a simple analytical understanding of them may not be easily achieved. A full numerical analysis will be given in the next section.

\subsection{BM case}
As discussed previously, without including RG corrections, there exist little allowed parameter space in the BM case for the low energy mixing angles in order to yield a prediction on $\delta$. Therefore, without RG corrections this BM scenario is already disfavored by the current data. An immediate question is whether it is possible to revive this BM case when the RG running effects are included. For this purpose, in this subsection, we discuss two examples: Case I with $\tan\beta = 30$ and $m^{}_0 = 0.005~{\rm eV}$ for negligible running effects, and Case II with $\tan\beta = 50$ and $m^{}_0 = 0.05~{\rm eV}$ for significant RG corrections.

\subsubsection*{Case I: BM, NO, $\tan\beta = 30$, $m^{}_0 = 0.005~{\rm eV}$}

As the absolute neutrino mass $m_0^{}$ is tiny, RG running effects are quite small for all three mixing angles and phases. Quantitatively, we can employ the RG equations of mixing angles and phases from Appendix A to estimate the RG corrections. First, given $y_\tau^{} \sim 0.3$, we have $y_\tau^2/(32\pi^2) \sim 0.3 \times 10^{-3}$. Then, with the logarithm of the ratio of two energy scales, i.e., $\ln (10^{12}/10^3) \sim 21$, one obtains $|\Delta\theta_{12}^{}| \lesssim 0.5^\circ$, $|\Delta\theta_{13}^{}| \lesssim 0.1^\circ$, $|\Delta\theta_{23}^{}| \lesssim 1^\circ$ and $|\Delta \delta| \lesssim 1^\circ$, which are negligible as we expect.

\begin{figure}
\centering
\includegraphics[scale=0.9]{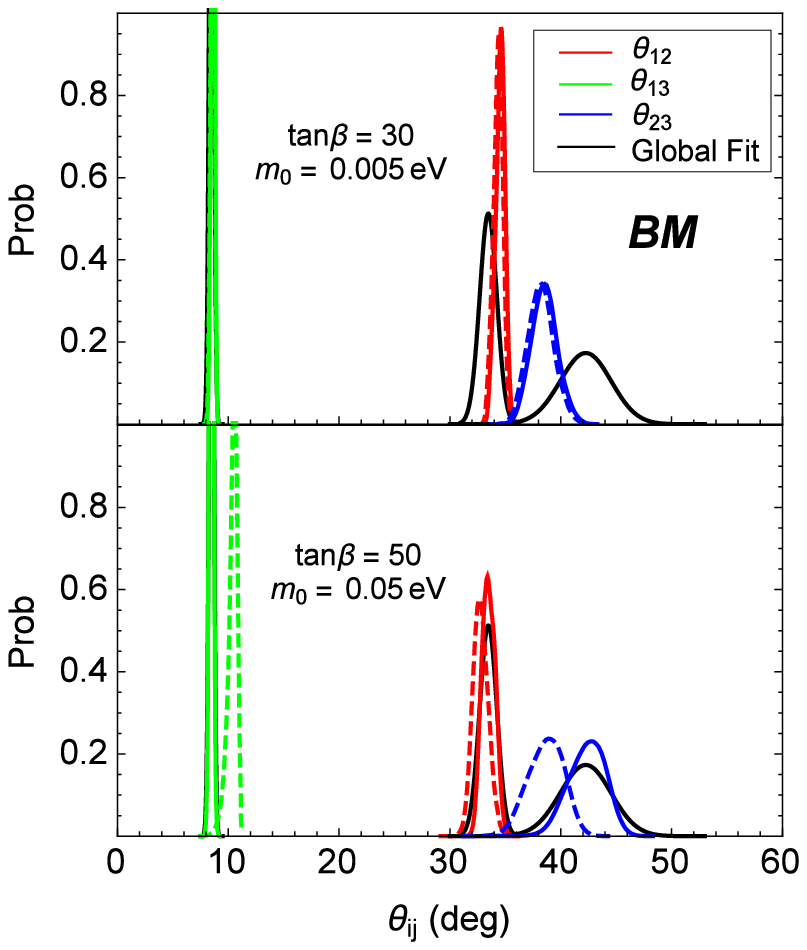}
\includegraphics[scale=0.93]{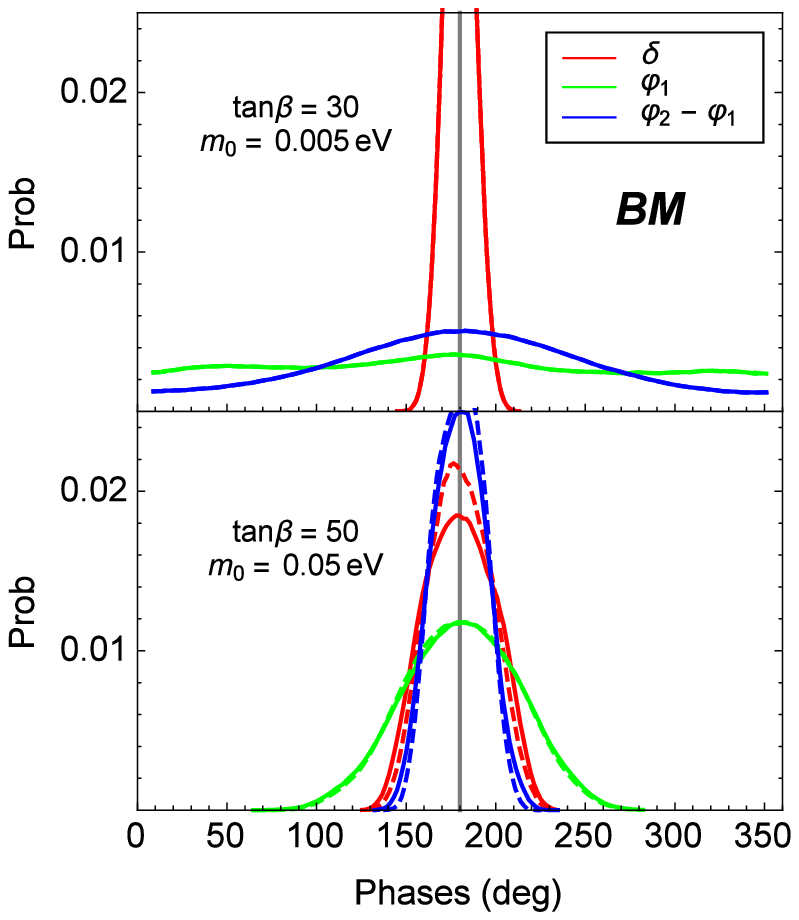}
\caption{Posterior distributions of neutrino mixing parameters at low and high energies in Case I (top) and II (bottom). On the left panel we present the results for the three mixing angles and their priors used in the numerical study, while on the right the results of three phases are shown. Solid and dashed curves with colors indicate low and high energy neutrino parameters, respectively.}
\label{fg:Case_I_II}
\end{figure}

On two top panels of Fig.~\ref{fg:Case_I_II} we show the posterior distributions of both low (solid curves) and high (dashed curves) energy neutrino parameters. On the left, the results for the three mixing angles (red, green and blue curves for $\theta_{12}^{}$, $\theta_{13}^{}$ and $\theta_{23}^{}$, respectively) are shown, together with their used priors (black curves) according to the recent global fit results. As one can see, the running of all the mixing angles and phases is indeed insignificant, with the solid and dashed curves almost indistinguishable.

Because of insufficient contributions from the RG running, this case is almost identical to that without RG corrections. Namely, satisfying the sum rule at high energy requires mixing angles at low energy to deviate from the regions favored by the recent global fit results. On the top-left panel of Fig.~\ref{fg:Case_I_II}, one can see that $\theta_{12}^{\rm L}$ (red and solid curve) resides at the large end of its prior distribution, and $\theta_{23}^{\rm L}$ (blue and solid curve) favors smaller values. With Eq.~(\ref{eq:SSR}) one can check that both of these two observations can help $\cos\delta^{\rm L}$ to obtain a more positive value. Finally, the predicted $\delta^{\rm L}$ (red and solid curve) on the top-right panel of Fig.~\ref{fg:Case_I_II} is still narrowly centered around $180^\circ$, as observed in the case without RG corrections~\cite{Ballett:2014dua}.

\subsubsection*{Case II: BM, NO, $\tan\beta = 50$, $m^{}_0 = 0.05~{\rm eV}$}

In a similar way, we can estimate the RG running to neutrino parameters, i.e., $|\Delta\theta_{13}^{}| \lesssim 2^\circ$ and $|\Delta\theta_{23}^{}| \lesssim 4^\circ$, while more than $10^\circ$ of running can be easily found for $\theta_{12}^{}$ and $\delta$. Due to this significant RG running, it is now possible the BM to reach a good agreement with low-energy neutrino oscillation data. This can be seen from the bottom-left panel of Fig.~\ref{fg:Case_I_II}, where the posterior distributions of three mixing angles at low energies are quite similar to their priors.

In addition, we also observe several other interesting features, which can be quickly explained. First, since the RG evolution of $\theta_{23}^{}$ is always in the negative direction at leading order, the high-energy value of $\theta_{23}^{}$ is smaller than its low-energy counterpart. In fact, smaller values at high energies are helpful to obtain a solution for $\delta$, as seen in the previous case. Second, in this quasi-degenerate scenario the difference between two Majorana phases has to be around $180^\circ$ so as to suppress the significant negative running of $\Delta$. This can be observed on the bottom-right panel of Fig.~\ref{fg:Case_I_II} (see blue curves). Moreover, due to this phase difference, the running of $\theta_{12}^{}$ is suppressed, as one can see from the bottom-left panel.
Third, we attempt to explain why the Majorana phase $\varphi_1^{}$ mostly sits in the second and third quadrants and why the high-energy value of $\theta_{13}^{}$ turns out to be larger than its low-energy value. To this end, we employ the expression of $\dot{\Delta}$ in Eq.(\ref{eq:dot_Delta_deg}) and insert $\varphi_2^{} -\varphi_1^{} \approx 180^\circ$. A straightforward calculation leads to
\begin{eqnarray}
\frac{32\pi^2}{y_\tau^2} \dot{\Delta} &\approx & \left \{ 2\sin^2(2\theta_{12}^{})\sin(2\theta_{23}^{})\cos\varphi_1^{} - 4 t_{23}^{} ( s_{12}^{\nu 2} - s_{12}^2) [1 - \cos\varphi_1^{} \cos(2\theta_{12}^{})] \right \}  \frac{m_0^2}{\Delta m_{32}^2} \; .
\end{eqnarray}
Therefore, in order to achieve a large and positive value of $\Delta$ at low energies, it is favorable to have $\dot{\Delta} < 0$, which leads to a constraint on $\varphi_1^{}$. As a rough estimation, using the best-fit values of three mixing angles, we find $69^\circ < \varphi_1^{} < 291^\circ$, which agrees with the observation that $\varphi_1^{}$ favors the second and third quadrants. Moreover, $\varphi_1^{} \sim 180^\circ$ minimizes the right-hand side of the above equation, resulting in the largest possibly positive contributions to $\Delta$. The above favored range of $\varphi_1^{}$ also explains why $\theta_{13}^{}$ tends to decrease when running towards low energies. More explicitly, the equation $\dot{\theta}_{13} \propto \cos(\varphi_1^{} - \delta)$ holds approximately, which tends to be positive when $\varphi_1^{}$ sits in the second and third quadrants and $\delta \sim 180^\circ$. Lastly, it is worthwhile to emphasize that because of significant RG running in this case, the predicted Dirac phase $\delta$ has a broader distribution around $180^\circ$, compared to the previous case.

\subsection{TBM case}
In the case of TBM, $\theta_{12}^\nu$ is very close to its measured value of $\theta^{}_{12}$ at low energies, so current neutrino oscillation data are already well compatible with the sum rule if no RG corrections are considered. However, this situation may be spoiled by RG corrections, as large RG running contributions to $\Delta$ could cause no solution for $\delta$ at low energies. As a result of this constraint, the neutrino parameters at high energies have to reside at some non-trivial ranges. In this subsection, we will also present two examples for TBM, Case III and Case IV, for which different patterns of mixing angles at high energies are observed.

\subsubsection*{Case III: TBM, NO, $\tan\beta = 30$, $m_0^{} = 0.05~{\rm eV}$}

Let us also first estimate the running effects of three mixing angles and the Dirac phase $\delta$. Similar inspections as before lead to $|\Delta\theta_{13}^{}| \lesssim 0.5^\circ$ and $|\Delta\theta_{23}^{}| \lesssim 1^\circ$, while more than $10^\circ$ of running can be easily found for $\theta_{12}^{}$ and $\delta$. Unlike Case II, where one needs to suppress the running of $\theta_{12}^{}$ by requiring $\varphi_2^{} -\varphi_1^{} \approx 180^\circ$ so that a better agreement with data at low energies can be obtained, we can now tolerate a large running of $\theta_{12}^{}$, as long as it is not too large.

\begin{figure}
\centering
\includegraphics[scale=0.9]{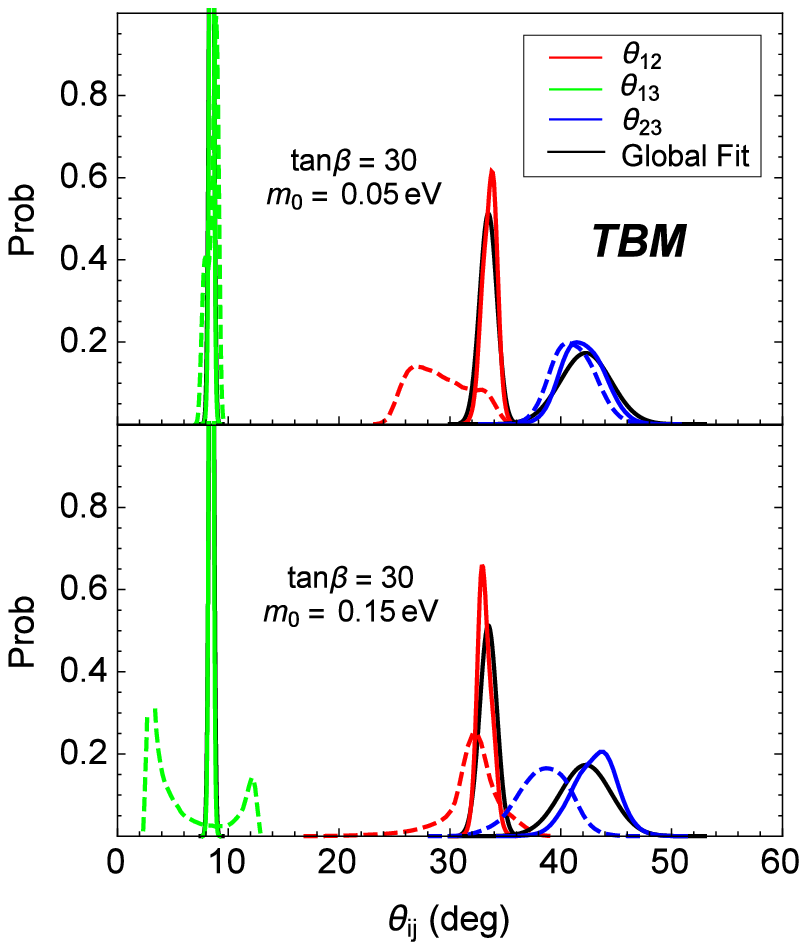}
\includegraphics[scale=0.93]{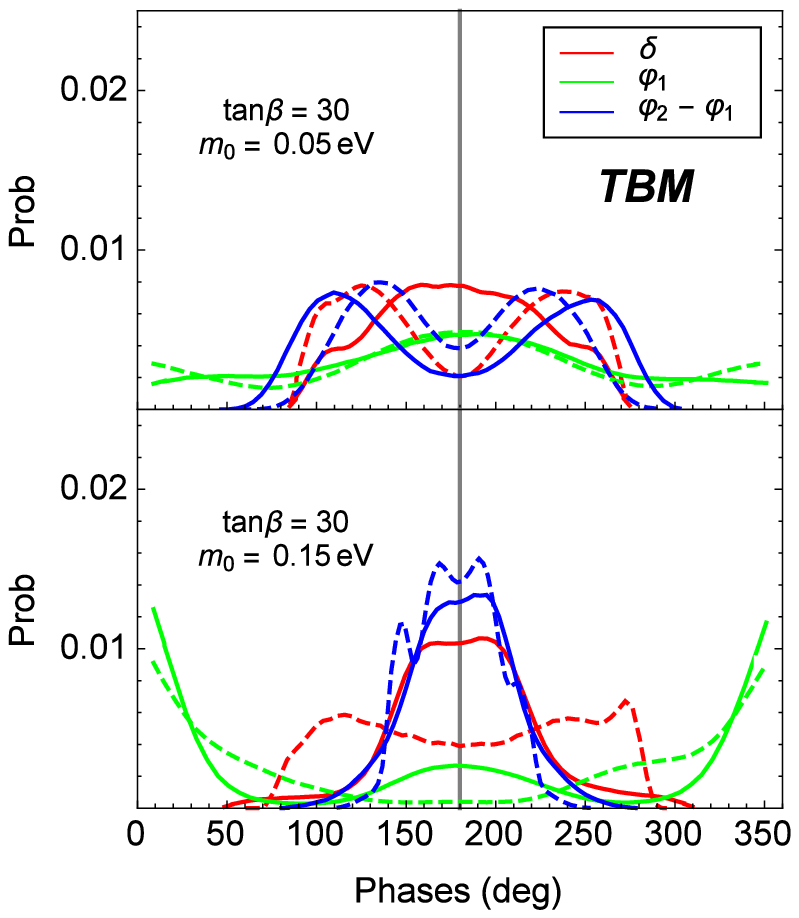}
\caption{Posterior distributions of neutrino mixing parameters at low and high energies in Case III (top) and IV (bottom). The rest of descriptions are the same as Fig.~\ref{fg:Case_I_II}.}
\label{fg:Case_III_IV}
\end{figure}

A simple way to figure out the allowed running of $\theta_{12}^{}$ is to take a close look at the sum rule at high energies. Assuming $\theta_{12}^{\rm H}$ at high energy to be not far away from $\theta_{12}^\nu$, we can use the sum rule at leading order, i.e., $\theta_{12}^{\rm H} \approx \theta_{12}^{\nu} + \theta_{13}^{\rm H} \cos\delta^{\rm H}$. Given $\theta_{13}^{\rm H} \sim \theta_{13}^{\rm L} \sim 9^\circ$, we roughly need $\theta_{12}^{\rm H} > 24^\circ$ to guarantee a solution of $\delta^{\rm H}$. Such a finding indeed agrees with the numerical result given in the top-left panel of Fig.~\ref{fg:Case_III_IV}.
The allowed running range of $\theta_{12}^{}$ also leads to a corresponding constraint on the difference between two Majorana phases $\varphi_2^{} - \varphi_1^{}$. With the help of the RG equation of $\theta_{12}^{}$, one can verify that requiring the running of $\theta_{12}^{}$ to be less than $10^\circ$ yields $ 115^\circ \lesssim \varphi_2^{} - \varphi_1^{} \lesssim 245^\circ$, which roughly agrees with the results given in the top-right panel of Fig.~\ref{fg:Case_III_IV}.

Finally, we point out that the allowed large running of $\theta_{12}^{}$ implies a positive $\dot{\Delta}$, and thus a negative $\Delta^{\rm L}$ at low energies. Consequently, the predicted $\delta^{\rm L}$ would favor the second and third quadrants, as we can see from Fig.~\ref{fg:Case_III_IV}.

\subsubsection*{Case IV: TBM, NO, $\tan\beta = 30$, $m_0^{} = 0.15~{\rm eV}$}

In Case IV with relatively large neutrino masses, RG running can be significant for all the mixing angles and phases. As an immediate consequence, according to Eq.~(\ref{eq:dot_Delta_deg}), $\Delta$ receives an enormous contribution from the last term that involves the difference between two Majorana phases $\varphi_2^{} - \varphi_1^{}$. To ensure a solution for $\delta^{\rm L}$ one has to suppress this contribution by requiring $\varphi_2^{} - \varphi_1^{} \sim 180^\circ$. Such a requirement also leads to an insignificant running of $\theta_{12}^{}$. Both of these findings are in a good agreement with the numerical results given in the bottom panels of Fig.~\ref{fg:Case_III_IV}.

Some further comments are in order. First, from the above observations of  $\varphi_2^{} - \varphi_1^{} \sim 180^\circ$ and a negligible running of $\theta_{12}^{}$, Eq.~(\ref{eq:dot_Delta_deg}) can be simplified to
\begin{eqnarray} \label{eq:Delta_CIV}
\frac{32\pi^2}{y_\tau^2} \dot{\Delta} &\approx & 2\sin^2(2\theta_{12}^{})\sin(2\theta_{23}^{}) \frac{m_0^2}{\Delta m_{32}^2} \cos\varphi_1^{} \; .
\end{eqnarray}
Integrating the above equation, we find that at low energies $\Delta^{\rm L} \sim - 0.1 \cos\varphi_1^{}$ and thus $\cos\delta^{\rm L} \sim - \cos\varphi_1$. Roughly speaking, the relations $(\varphi_1 \pm \delta) \sim 180^\circ$ seem to be favored, and they are consistent with the results given in the bottom-right panel of Fig.~\ref{fg:Case_III_IV}. Second, with $(\varphi_1 \pm \delta) \sim 180^\circ$, the peculiar double-peak distribution of $\theta_{13}^{}$ can be explained. To see this, we first notice that in the nearly-degenerate mass region the RG equation of $\theta_{13}^{}$ reduces to $\dot{\theta}_{13} \propto \cos(\varphi_1 -\delta)$. Thus, for the case $(\varphi_1 - \delta) \sim 180^\circ$ one has a larger value of $\theta_{13}^{}$ at low energies, implying the left peak. Regarding the right peak, it is due to the other possibility $(\varphi_1 + \delta) \sim 180^\circ$, in which $|\varphi_1^{} - \delta| < \pi/2$ occurs quite likely. Lastly, according to Eq.~(\ref{eq:Delta_CIV}) and the fact that $\varphi_1^{}$ favors the first and fourth quadrants leads to a more negative value of $\Delta^{\rm L}$, the predicted $\delta^{\rm L}$ tends to be in the second and third quadrants.

Before turning to the full numerical analysis, let us recapitulate what we have learned from the above detailed investigation of four examples:
\begin{itemize}
\item For the BM scenario, it is possible to have three mixing angles at low energies compatible with the latest global-fit results, as long as the RG corrections are significant enough.

\item In contrast, for the TBM scenario, the role played by the RG running effects is then to single out different favored distributions of the Dirac and Majorana phases, depending on the size of RG corrections. Moreover, because of those requirements on phases, flavor mixing angles show different patterns at high energies, which may provide some clues for the flavor model building.
\end{itemize}
In the next section, a few more scenarios will be considered, and a model comparison among them will also be performed.

\section{Full numerical results}

\begin{figure}[ht]
\centering
\includegraphics[scale=0.73]{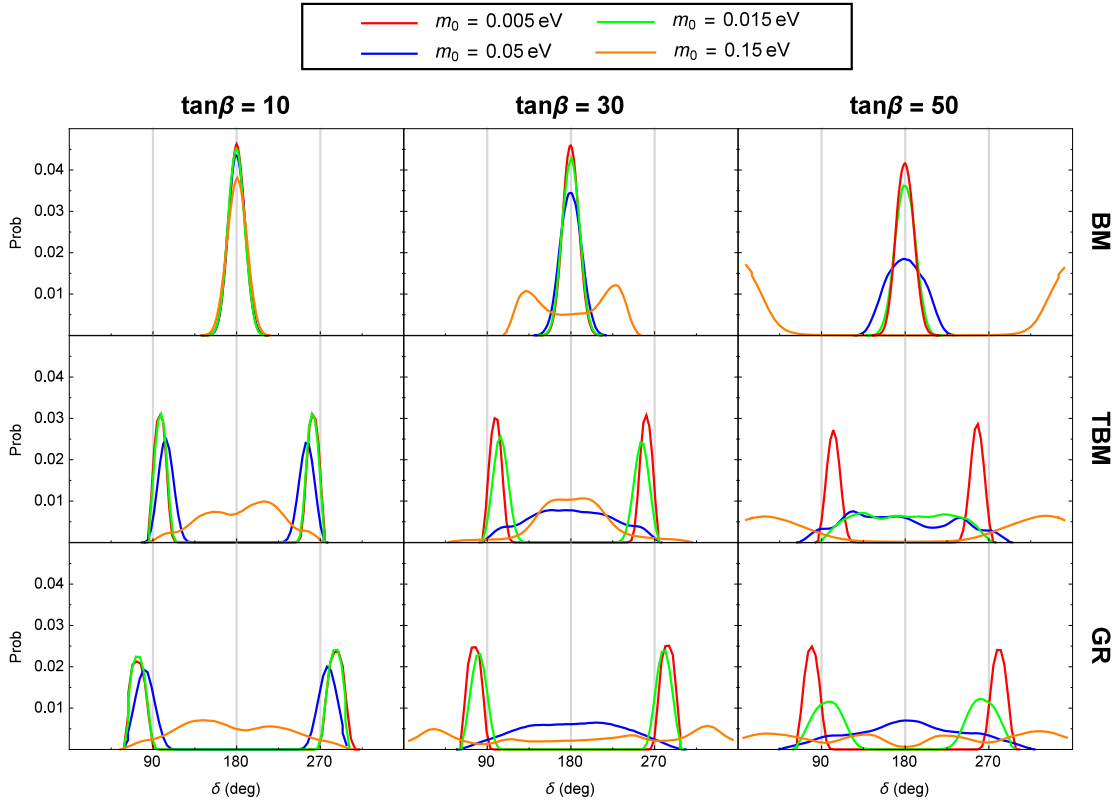}
\caption{Posterior distributions of the Dirac phase $\delta^{\rm L}$ at low energies in the scenario of NO. }
\label{fg:SSR_NH_global}
\end{figure}

\begin{figure}[ht]
\centering
\includegraphics[scale=0.73]{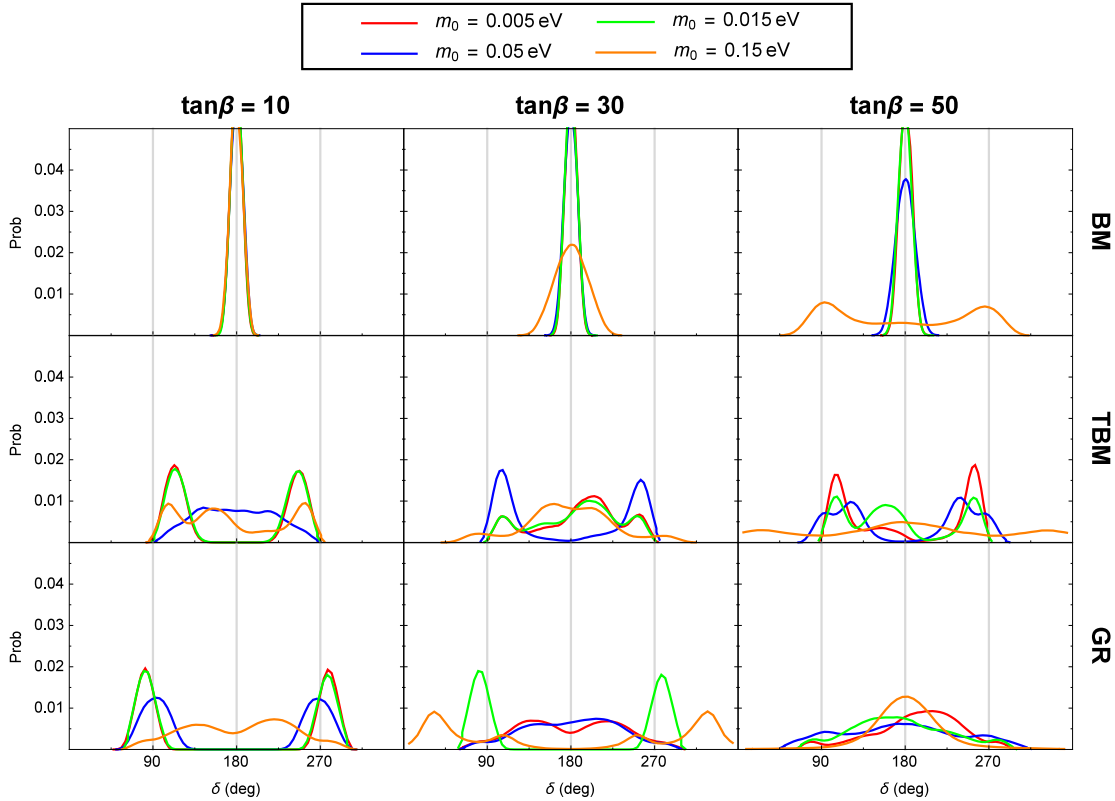}
\caption{Posterior distributions of the Dirac phase $\delta^{\rm L}$ at low energies in the scenario of IO. }
\label{fg:SSR_IH_global}
\end{figure}

We now present a complete numerical analysis of all the scenarios, which are labeled by three different constant mixing patterns (i.e., BM, TBM and GR), three choices of $\tan\beta$ (i.e., $\tan\beta = 10, 30, 50$), four different values of the lightest neutrino mass (i.e., $m_0^{} = 0.005~{\rm eV}$, $0.015~{\rm eV}$, $0.05~{\rm eV}$ and $0.15~{\rm eV}$), and a further distinguish of two neutrino mass orderings (i.e., NO or IO). While in our numerical study the posterior distributions of all neutrino parameters can be obtained, we restrict ourselves to only four of them, i.e., three mixing angles and the Dirac phase, as a precise determination of two Majorana phases will not be experimentally achievable in a near future. Moreover, among these four parameters, from the previous study we notice that satisfying the current neutrino data of mixing angles is not difficult for most cases, their posterior distributions thus almost follow the priors. In other words, imposing the solar mixing sum rule at high energies yields essentially no constraints on the mixing angles. Hence, in this section we focus on the posterior distributions of the Dirac phase $\delta^{\rm L}_{}$ at low energies, which are shown in Fig.~\ref{fg:SSR_NH_global} and Fig.~\ref{fg:SSR_IH_global} for NO and IO, respectively.

First, we study the case of NO by carefully examing the results given in Fig.~\ref{fg:SSR_NH_global}. The main observations can be summarized as follows:
\begin{itemize}
\item In all three mixing scenarios, increasing the strength of RG running via either a larger value of $m_0^{}$ or $\tan\beta$ yields significant distortions to the results in the case of no RG corrections, which can be approximately represented by $\tan\beta = 10$ and $m_0^{} = 0.005~{\rm eV}$. Such a distortion can be a much broader distribution, or a shift of peaks, or a disappearance of peaks, or a combination of them.

\item In the case of BM, one may neglect the impact from RG running when $\tan\beta \lesssim 30$ and $m_0^{} \lesssim 0.05~{\rm eV}$, while for TBM and GR, a stricter requirement is needed, i.e., $m_0^{} \lesssim 0.015~{\rm eV}$ for $\tan\beta \sim 30$.

\item When RG running effects are moderate, in the case of BM, $\delta^{\rm L}$ tends to favor regions that are away from $180^\circ$. On the contrary, for TBM and GR, moderate RG corrections would lead $\delta^{\rm L}$ to move towards $180^\circ$, as also discussed in the previous section. As a result, one may have significant overlap in the favored regions of $\delta^{\rm L}$ among all these three mixing scenarios, e.g., in the case of $\tan\beta \sim 30$ and $m_0^{} \sim 0.05~{\rm eV}$. Therefore, compared to the non-RG case, distinguishing BM from TBM and GR by a precise measurement of $\delta$ would be more difficult.

\item Finally, we also observe that when the RG running effects are very strong, i.e., $\tan\beta \gtrsim 30$ and $m_0^{} \gtrsim 0.15~{\rm eV}$,   the resulting prediction on $\delta^{\rm L}$ becomes less tractable, so that an experimental verification of solar mixing sum rule becomes less promising.

\end{itemize}

Then, we turn to the scenario of IO. In Fig.~\ref{fg:SSR_IH_global}, we observe that for the BM case, the change to the posterior distribution of $\delta^{\rm L}$ seems to follow the same trend as that observed in the NO case, i.e., becoming broader and starting to favor regions away from $180^\circ$. However, for the TBM and GR cases, the RG running effects on the prediction of $\delta^{\rm L}$ are very violent, and no characteristic feature can be easily identified.  Moreover, in this case we may only neglect the RG corrections to the solar mixing sum rule when $\tan\beta \lesssim 10$ and $m_0^{} \lesssim 0.015~{\rm eV}$.

Lastly, since we have adopted the Bayesian analysis to perform the numerical study, and three different hypotheses corresponding to those three different mixing scenarios have been considered, one then may be curious about which hypothesis is most compatible with current neutrino data, given the requirement of satisfying the solar mixing sum rule at high energies. In the previous discussion of four examples we have already opposed the BM case to the TBM case, but only in a qualitative manner, now we want to study this question more quantitatively, within the formalism of Bayesian analysis.

In Bayesian analysis, the comparison of degrees of belief of different hypotheses can be carried out by computing the ratios of so-called posterior odds between any two competing hypotheses. In our case, this posterior odds is simply ${\rm Pr}(\mathcal{H}_i^{})/{\rm Pr}(\mathcal{H}_j^{})$, which coincides with the Bayes factor $\mathcal{B}$, assuming equal prior probabilities for all hypotheses. Moreover, to interpret the value of this posterior odds or the Bayes factor, one often adopts the Kass-Raftery~\cite{KR} or Jeffreys~\cite{Jeffreys,Jeffreys_mod,Trotta:2008qt} scale. In Table~\ref{tb:Jeffreys} we list the Jeffreys scale that used in \cite{Jeffreys_mod,Trotta:2008qt}, and will implement them to interpret our numerical results.

\begin{table}
\centering
\begin{tabular}{l | l | l | l}
\hline
\hline
$\left|\ln(\text{odds})\right|$ & Odds & Probability & Interpretation \\
\hline
$< 1.0$ & $\lesssim 3 : 1$ & $\lesssim 75.0\%$ & Inconclusive \\
$1.0$ & $\simeq 3 : 1$ & $\simeq 75.0\%$ & Weak evidence\\
$2.5$ & $\simeq 12 : 1$ & $\simeq 92.3\%$ & Moderate evidence \\
$5.0$ & $\simeq 150 : 1$ & $\simeq 99.3\%$ & Strong evidence\\
\hline
\hline
\end{tabular}
\vspace{0.3cm}
\caption{The Jeffreys scale used for the statistical interpretation of Bayes factors, posterior odds and model probabilities~\cite{Jeffreys_mod, Trotta:2008qt}.}
\label{tb:Jeffreys}
\end{table}

\begin{figure}
\centering
\includegraphics[scale=0.55]{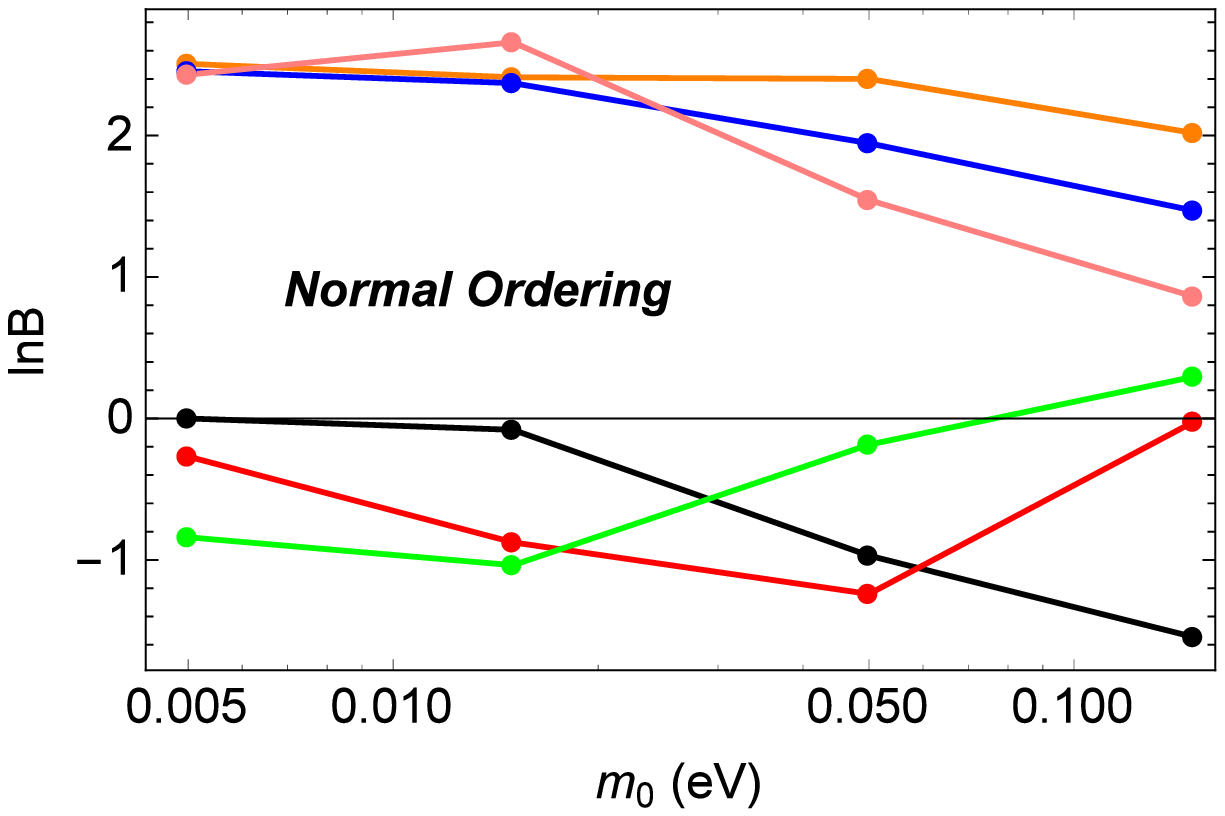}
\includegraphics[scale=0.77]{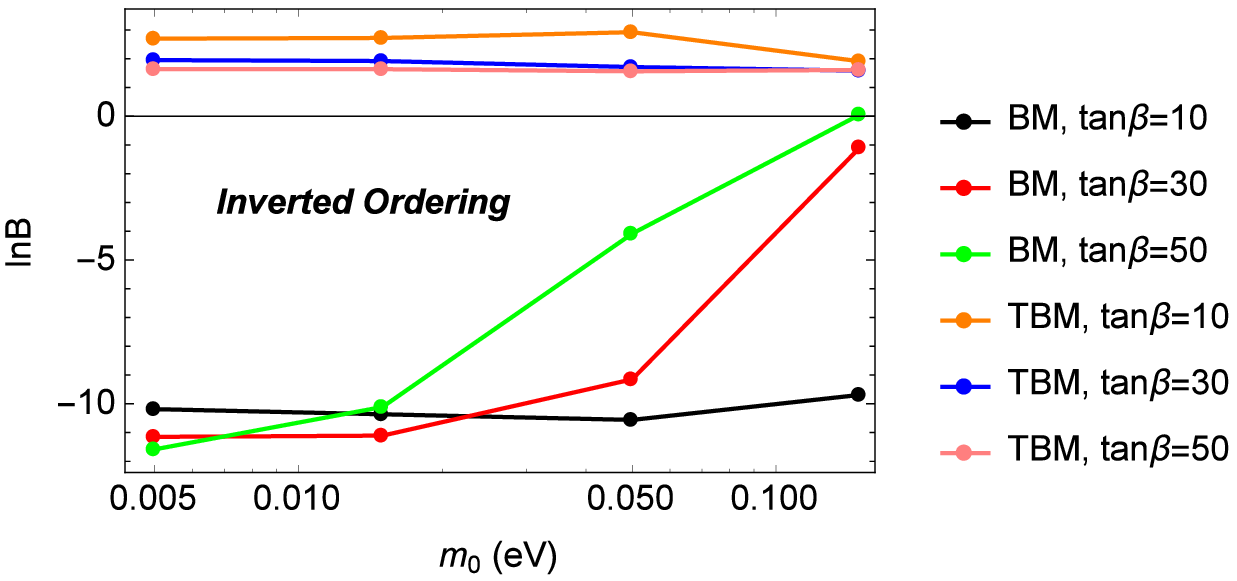}
\caption{Logarithm of the Bayes factors for all BM and TBM cases in both NO and IO scenarios. Here we choose the case with $m_0^{} = 0.005~{\rm eV}$, $\tan\beta = 10$, BM and NO as the benchmark case. Bayes factors of other cases are obtained by computing the ratios of their ${\rm Pr}(\mathcal{H})$ to that of the benchmark case.}
\label{fg:evidence}
\end{figure}

Thanks to the M{\scriptsize ULTI}N{\scriptsize EST} program we are able to calculate ${\rm Pr}(\mathcal{H})$ for all the cases under consideration. Assuming equal prior probabilities, we then can find out which one is more favorable by computing the Bayes factor. Choosing the benchmark case as the one with $m_0^{} = 0.005~{\rm eV}$, $\tan\beta = 10$, BM and NO, we plot the logarithm of the Bayes factors, $\ln \mathcal{B}$, for all BM and TBM cases in both NO and IO scenarios in Fig.~\ref{fg:evidence}. The results for the GR case are not shown here, as they are quite similar to those for the TBM.

In Fig.~\ref{fg:evidence}, we observe that in the scenario of NO, when the RG running effects are insignificant, i.e., $\tan\beta \lesssim 30$ and $m_0^{} \lesssim 0.015~{\rm eV}$, the differences of $\ln\mathcal{B}$ between the TBM and BM cases are around 2.5, indicating a moderate preference for TBM according to the Jeffreys scale in Table~\ref{tb:Jeffreys}. However, when RG effects becomes non-negligible, the preference between the two cases gets diminished, e.g., the difference of $\ln\mathcal{B}$ can even reduce to less than one for the case with $\tan\beta = 50$ and $m_0^{} = 0.15~{\rm eV}$. For the IO scenario, the above finding not only holds but also becomes more evident. Hence, we can conclude that while in general the TBM (or GR) case is more compatible with current neutrino data as opposed to the BM case, RG running effects can weaken such a preference, so that singling out one promising scenario among others becomes more difficult.

\section{Summary and conclusions}

Solar mixing sum rules are generic predictions that relate the lepton mixing angles and the Dirac CP-violating phase. They arise when the corrections from the charged-lepton sector are added to the mixing matrix in the neutrino sector, with the assumption that the latter takes particular forms of constant mixing matrices, such as Bimaximal, Tri-Bimaximal and Golden Ratio. Motivated by the fact that these sum rules can be derived at high energies, we set out in this work to study the RG running effects on them, and pay particular attention to their predictions for the Dirac CP-violating phase at low energies. For illustration, we choose to work within the framework of MSSM, where large RG running effects can be present if $\tan\beta$ is relatively large.

To quantify the RG corrections to the sum rule, we have introduced a parameter $\Delta$, which vanishes at the high-energy scale and deviates from zero at the low-energy scale.\footnote{In this connection, it is worth mentioning that the radiative corrections to a possible $\mu$-$\tau$ symmetry in the MNSP matrix~\cite{Xing:2015fdg} have already been investigated in Ref.~\cite{Luo:2014upa}.} It is found that the size of such a deviation is related to the running of the lepton flavor mixing angles and the Dirac phase, and too large running of these neutrino parameters may lead to no solution for the Dirac phase at low energy. We have carefully studied this impact on the low-energy lepton mixing parameters in four special cases, and find that when RG effects are too large, the two Majorana phases need to differ by around $180^\circ$ so as to suppress the running of mixing angles and phases.

In a full numerical analysis, we adopt the notion of Bayesian statistical approach. We choose the lepton mixing parameters at low energies as our parameters of interest. Current global-fit results are used for their prior distributions at low energies, and then the impact of imposing solar mixing sum rule at a high-energy scale can be addressed by analyzing the posterior distributions. To relate parameters at low- and high-energy boundaries, we solve the RG equations of all lepton mixing parameters numerically. Our main conclusions are summarized below:
\begin{itemize}
\item First, in the case of BM, RG corrections improve the agreement with current neutrino data for the three mixing angles. Without RG corrections, satisfying the solar mixing sum rule would require the three mixing angles to lie in the regions that are less favored by data. Moreover, with more significant running effects, the predicted Dirac phase $\delta$ have a broader distribution around $180^\circ$, indicating that an experimental verification is more difficult. In addition, when RG effects are too large, the favored region can be quite far away from $180^\circ$.

\item Second, for the cases of TBM and GR, the role played by the RG running effects is to single out different favored distributions of the Dirac and Majorana phases, resulting in different patterns of mixing angles and phases at high energies. Regarding the prediction for $\delta$, as in the BM case, including RG running effects would make the distribution broader on the one hand, and the favored regions tend to be closer to $180^\circ$ on the other hand, if RG running is moderate. However, when the RG running effects are violent, especially in the scenario of inverted neutrino mass ordering, the predicted distributions of $\delta$ are rather complicated and no obvious features can be identified.

\item Third, we investigate the possibility to discriminate the case of BM from the TBM and GR cases in the presence of RG running. It is found that RG corrections could lead to a remarkable overlap in the favored regions of the Dirac phase among all three cases. As a result, a discrimination of them by a precise measurement of the Dirac phase seems less promising. Moreover, we also find that RG running effects can weaken the preference of TBM and GR over BM.
\end{itemize}
It should be noticed that the above conclusions are restricted to the scenario of MSSM, and a separate but similar analysis to the one given here needs to be performed for other extensions of the SM. However, we have demonstrated that in order to explore possible sum rules or flavor symmetries at high energies, one has to find out whether the RG running effects are small or not, in addition to precise measurements of neutrino mixing parameters at low energies. In MSSM, this can be done by pinning down neutrino mass ordering, the absolute neutrino mass scale and the value of $\tan\beta$. The further extension of current work to other sum rules of leptonic mixing and to a different framework is also interesting and deserves a dedicated study.

\hspace{0.2cm}
\begin{flushleft}
{\bf Acknowledgements}
\end{flushleft}
This work was supported in part by the National Recruitment Program for Young Professionals and by the CAS Center for Excellence in Particle Physics (CCEPP).

\appendix

\section{RG equations of lepton mixing parameters}

In this appendix, we list the RG equations of three mixing angles and three CP-violating phases, essentially following the notations of Ref.~\cite{Antusch:2003kp}. The RG equations of three mixing angles are given by
\begin{eqnarray}
\frac{\text{d} \theta_{12}}{\text{d} t} &=& -\frac{y_\tau^2}{32\pi^2}\sin2\theta_{12} s_{23}^2 \frac{|m_1 e^{i\varphi_1} + m_2 e^{i\varphi_2}|^2}{\Delta m_{21}^2} + \mathcal{O}(\theta_{13}) \;, \\
\label{eq:RGE_t13}
\frac{\text{d}\theta_{13}}{\text{d} t} &=& \frac{y_\tau^2}{32\pi^2} \sin 2\theta_{12} \sin 2\theta_{23} \frac{m_3}{\Delta m_{32}^2(1+\zeta)} \times \nonumber \\
& & \times [m_1 \cos(\varphi_1 - \delta) - (1+\zeta) m_2 \cos(\varphi_2 - \delta) - \zeta m_3 \cos\delta] + \mathcal{O}(\theta_{13}) \; , \\
\frac{\text{d}\theta_{23}}{\text{d} t} &=& -\frac{y_\tau^2}{32\pi^2} \sin 2\theta_{23} \frac{1}{\Delta m_{32}^2} \left [ c_{12}^2 |m_2e^{i\varphi_2} + m_3|^2 + s_{12}^2 \frac{|m_1 e^{i\varphi_1}+m_3|^2}{1+\zeta} \right ] + \mathcal{O}(\theta_{13}) \; ,
\end{eqnarray}
where $\Delta m_{21}^2 = m_2^2 - m_1^2$ and $\Delta m_{32}^2 = m_3^2 - m_2^2$, and their ratio is defined as $\zeta = \Delta m_{21}^2/\Delta m_{32}^2$. The RG equation of the Dirac phase reads
\begin{eqnarray}
\frac{\text{d} \delta}{\text{d} t} &=& \frac{y_\tau^2}{32\pi^2} \frac{\delta^{(-1)}}{\theta_{13}} + \frac{y_\tau^2}{8\pi^2} \delta^{(0)} + \mathcal{O}(\theta_{13}) \; ,
\end{eqnarray}
where
\begin{eqnarray}
\delta^{(-1)} &=& \sin 2\theta_{12} \sin 2\theta_{23} \frac{m_3}{\Delta m_{32}^2 (1+\zeta)} \times \nonumber \\
& &~\times [m_1 \sin(\varphi_1 - \delta) - (1+\zeta) m_2 \sin (\varphi_2 -\delta) + \zeta m_3 \sin \delta] \; , \\
\delta^{(0)} &=& \frac{m_1 m_2 s_{23}^2 \sin(\varphi_1 - \varphi_2)}{\Delta m_{21}^2} \nonumber \\
& &~ + m_3 s_{12}^2 \left [ \frac{m_1 \cos 2\theta_{23} \sin\varphi_1}{\Delta m_{32}^2	(1+ \zeta)} + \frac{m_2 c_{23}^2 \sin(2\delta-\varphi_2)}{\Delta m_{32}^2} \right ] \nonumber \\
& &~ + m_3 c_{12}^2 \left [ \frac{m_2 \cos 2\theta_{23} \sin\varphi_2}{\Delta m_{32}^2	} + \frac{m_1 c_{23}^2 \sin(2\delta-\varphi_1)}{\Delta m_{32}^2(1+\zeta)}  \right ] \; ;
\end{eqnarray}
and those for the Majorana CP phases
\begin{eqnarray}
\frac{\text{d} \varphi_1}{\text{d} t} &=& \frac{y_\tau^2}{4\pi^2} \left \{ m_3 \cos 2\theta_{23} \frac{m_1 s_{12}^2 \sin \varphi_1 + (1+\zeta) m_2 c_{12}^2 \sin \varphi_2}{\Delta m_{32}^2 (1+\zeta)} \right. \nonumber \\
& & ~~~~\left. + \frac{m_1 m_2 c_{12}^2 s_{23}^2 \sin(\varphi_1 - \varphi_2)}{\Delta m_{21}^2} \right \} + \mathcal{O}(\theta_{13}) \; , \\
\frac{\text{d} \varphi_2}{\text{d} t} &=& \frac{y_\tau^2}{4\pi^2} \left \{ m_3 \cos 2\theta_{23} \frac{m_1 s_{12}^2 \sin \varphi_1 + (1+\zeta) m_2 c_{12}^2 \sin \varphi_2}{\Delta m_{32}^2 (1+\zeta)} \right. \nonumber \\
& & ~~~~\left. + \frac{m_1 m_2 s_{12}^2 s_{23}^2 \sin(\varphi_1 - \varphi_2)}{\Delta m_{21}^2} \right \} + \mathcal{O}(\theta_{13}) \; .
\end{eqnarray}

\end{document}